\documentclass[aps,pre,twocolumn,showpacs,amsmath,amssymb]{revtex4}
\usepackage{graphicx}
\usepackage{epsf}
\usepackage{bm}

\newcommand{\bi}[1]{\mbox{\boldmath$#1$}}

\begin{document}

\title{
Multiple time scales hidden in heterogeneous dynamics of glass-forming liquids
}

\author{Kang Kim}
\affiliation{
Institute for Molecular Science, Okazaki 444-8585, Japan
}

\author{Shinji Saito}
\affiliation{
Institute for Molecular Science, Okazaki 444-8585, Japan
}

\begin{abstract}
A multi-time probing of density fluctuations is introduced to
 investigate hidden time scales of heterogeneous dynamics in
 glass-forming liquids.
Molecular dynamics simulations for simple glass-forming
 liquids are performed, and a three-time correlation function is numerically
 calculated for general time intervals.
It is demonstrated that the three-time correlation function is sensitive to the
 heterogeneous dynamics and that it reveals couplings of correlated motions
 over a wide range of time scales.
Furthermore, the time scale of the heterogeneous dynamics $\tau_{\rm hetero}$ is
 determined by the change in the second time interval in the three-time correlation
 function.
The present results show that the time scale of the heterogeneous dynamics
 $\tau_{\rm hetero}$ becomes larger than the $\alpha$-relaxation time at
 low temperatures and large wavelengths.
We also find a dynamical scaling relation between the time scale
 $\tau_{\rm hetero}$ and the length scale $\xi$ of dynamical
 heterogeneity as  $\tau_{\rm hetero} \sim \xi^{z}$ with $z=3$.
\end{abstract}

\date{\today}

\maketitle


The understanding of drastic slowing down and non-exponential
relaxations in
glasses and dense colloids and the jamming of granular materials
is one of the most challenging problems in condensed matter physics.
Extensive studies have been carried out through experiments, computer
simulations, and theories~\cite{Binder2005Glassy, Liu2001Jamming}.
Recently, the concept of ``dynamical heterogeneity'' in 
glass-forming liquids has attracted much attention
and has been considered as an essential notion for understanding the
slow dynamics.
Near the glass transition temperature, the dynamics becomes spatially
heterogeneous and include the coexistence of mobile and
immobile correlated regions.
A number of molecular
dynamics simulations~\cite{Hurley1995Kinetic, Kob1997Dynamical,
Donati1998Stringlike, Muranaka1995Beta, Yamamoto1998Dynamics,
Yamamoto1998Heterogeneous, Perera1999Relaxation, Doliwa2002How,
Glotzer2000Spatially, Lacevic2003Spatially, Berthier2004Time} and
experiments~\cite{Kegel2000Direct, Weeks2000Threedimentional}
have detected the existence of dynamical
heterogeneity with various visualizations.

To characterize the heterogeneous dynamics in glass-forming
liquids,
explorations that can provide more detailed information that is not
available from conventional two-point correlation functions,
are essential and significant.
In fact, the growth of the length scale $\xi$ with decreasing temperature has been
measured in terms of four-point correlation
functions by simulations 
~\cite{Yamamoto1998Dynamics, Yamamoto1998Heterogeneous, Perera1999Relaxation, 
Glotzer2000Spatially, Lacevic2003Spatially,
Berthier2004Time, Toninelli2005Dynamical, Chandler2006Lengthscale,
Richard2008Scaling}, experiments~\cite{Ediger2000Spatially,
Richert2002Heterogeneous, Berthier2005Direct}, and mode-coupling
theory~\cite{Biroli2006Inhomogeneous}.

Another important issue is the temporal
details of the dynamical heterogeneity, such as the lifetime and the relocation
time, i.e., 
the time scale over which
slow (fast) moving particles remain slow (fast).
The central question is whether or not the lifetime of dynamical heterogeneity
$\tau_{\rm hetero}$ is
comparable to the so-called $\alpha$-relaxation time $\tau_\alpha$
determined by the two-point correlation function~\cite{Ediger2000Spatially,
Richert2002Heterogeneous}.
If the deviation between the two time scales becomes large near the glass
transition temperature,
the details of the relaxation processes
of spatially heterogeneous dynamics 
are essential to understanding the drastic slowing down.

In order to quantify $\tau_{\rm hetero}$, a multiple time extension of the density
correlation function is necessary, since the two-point correlation
function averaging over the full ensemble cannot distinguish among 
dynamics of sub-ensembles.
This idea has been applied to 
various experiments such as multidimensional nuclear magnetic
resonance, hole-burning, and photobleaching~\cite{Schmidt1991Nature,
Bohmer1996Dynamic, Wang1999How, Wang2000Lifetime,
Ediger2000Spatially, Richert2002Heterogeneous}.
Recently, several experiments have provided the evidence to support that the
lifetime of heterogeneous dynamics $\tau_{\rm hetero}$ becomes larger than the 
structural relaxation time near the glass
transition~\cite{Wang1999How, Wang2000Lifetime}.
In computer simulations, the multiple time extension has also been employed
to determine the lifetime of heterogeneous
dynamics~\cite{Heuer1997Heterogeneous, Heuer1997Information, Yamamoto1998Heterogeneous,
Doliwa1998Cage, Perera1999Relaxation, Doliwa2002How,
Flenner2004Lifetime, Leonard2005Lifetime, Jung2005Dynamical}.
In these calculations, however, due to intense calculations several time
intervals are fixed at a characteristic time scale,
and thus only limited information on the lifetime of the heterogeneous
dynamics has been provided.


In this paper, we present
the comprehensive information regarding
the lifetime of dynamical heterogeneity and its temperature dependence.
We perform molecular dynamics (MD)
simulations for simple glass-forming liquids at various temperatures
and investigate the dynamics in terms of a multi-time correlation
function that shows the coupling of motions with various time scales in
the heterogeneous dynamics.
The wave vector dependence of the lifetime of heterogeneous
dynamics is also studied.
Furthermore, the dynamical scaling relation between the
lifetime and the length scale of the dynamical heterogeneity is presented.

\begin{figure}[t]
\includegraphics[width=.4\textwidth]{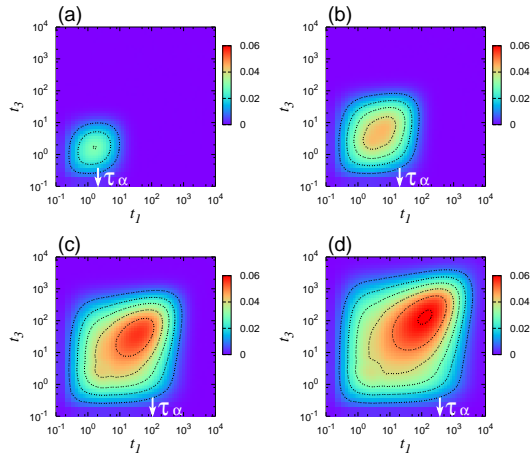}
\vspace*{-0.5cm}
\caption{Two-dimensional plots of the three-time correlation
 function $F_4(k, t_1, t_2, t_3)$
 of component $1$ particles
at waiting time $t_2=0$ for various temperatures $T=0.473$ (a), $0.352$
(b), $0.306$ (c), and $0.289$ (d).
For each temperature, $\tau_\alpha$ is indicated by the arrow.
The wave vector $k$ is chosen as $k=2\pi$.}
\vspace*{-0.5cm}
\label{F4_0t2}
\end{figure}

Similar to earlier studies~\cite{Heuer1997Heterogeneous,
Heuer1997Information, Doliwa1998Cage, Flenner2004Lifetime,
Leonard2005Lifetime}, 
we define the three-time, i.e., four-point, correlation function
with times $0$, $\tau_1$, $\tau_2$, and $\tau_3$,
\begin{equation}
F_4(k, t_1, t_2, t_3) = \biggl\langle  
\frac{1}{N}\sum_{j=1}^N  \delta F_j(\bi{k}, \tau_2, \tau_3)
\delta F_j(\bm{k}, 0, \tau_1))\biggr\rangle,
\label{three_time_correlation}
\end{equation}
where
$
\delta F_j(\bi{k}, 0, \tau) = \cos[\bi{k}\cdot\Delta \bi{r}_j(0, \tau)]-F_s(k, \tau)
$
is the individual fluctuation in the incoherent intermediate
scattering function defined as
$F_s(k, \tau)=\langle (1/N) \sum_{j=1}^N \cos(\bi{k}\cdot\Delta \bi{r}_j(0, \tau))\rangle$
with the wave vector $\bi{k}$ and $k=|\bi{k}|$.
Here,
$\Delta{\bi r}_j(0, \tau)={\bi r}_j(\tau)-{\bi r}_j(0)$ is 
the displacement vector between two times, $0$ and $\tau$ of $j$-th particle.
The $\langle \cdots \rangle$ 
represents the ensemble average over the initial time $t=0$ and the angular
components of the wave vector.
The definition of the time interval $t_i$ is $t_i=\tau_i - \tau_{i-1}$,
where $\tau_0=0$.
It is noted that $F_4(k, t_1, t_2, t_3)$ expresses
the correlations between fluctuations
in the two-point correlation function $F_s(k, t)$ between two time intervals, $t_1=\tau_1$
and $t_3=\tau_3-\tau_2$.
If the dynamics are homogeneous and if the motions between the two
intervals $t_1$ and $t_3$ are uncorrelated and decoupled, the multi-time
correlation function $F_4(k, t_1, t_2, t_3)$ should become zero.
On the other hand, if the dynamics are heterogeneous, spatially 
heterogeneous structures
governed by the dichotomy between fast- and slow-moving regions lead
to finite values of $F_4(k, t_1, t_2, t_3)$ due to correlations of
motions between the two intervals $t_1$ and $t_3$.
Furthermore, 
the progressive change in the waiting time $t_2=\tau_3-\tau_2$ of $F_4(k, t_1, t_2, t_3)$
enables us to 
investigate the time scale of the heterogeneous structure
in glass-forming liquids.
In other words, the first two-point function $\delta F_j(\bi{k}, 0,
\tau_1)$ is capable of selecting the sub-ensemble of slow (fast)
contributions in heterogeneities and 
the total function $\delta F_j(\bi{k}, 0,
\tau_1)\delta F_j(\bi{k}, \tau_2, \tau_3)$ can reveal how long
the difference in the dynamics of the sub-ensembles remains over the waiting
time $t_2$~\cite{Heuer1997Information}.
It should be noted that multi-time correlation functions are exploited
in nonlinear multidimensional spectroscopic studies of liquids and proteins to
understand the detailed dynamics, e.g., the transition from
inhomogeneous to homogeneous dynamics and the couplings of molecular
motions~\cite{Mukamel1999Principles, Hochstrasser2007Twodimensional}.


\begin{figure}[t]
\includegraphics[width=.4\textwidth]{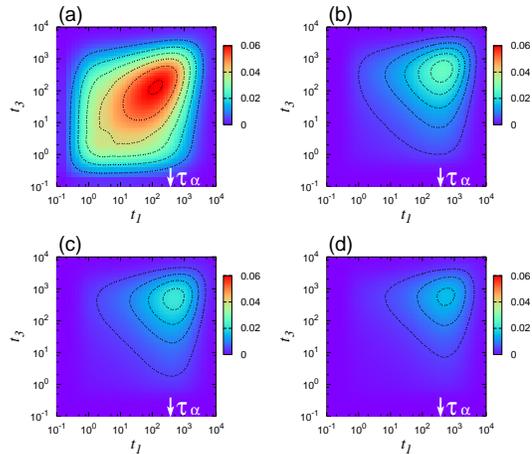}
\vspace*{-0.5cm}
\caption{Two-dimensional plots of the three-time correlation
 function $F_4(k, t_1, t_2, t_3)$
 of component $1$ particles
for $T=0.289$ at several waiting time $t_2=0$ (a), $0.5\tau_\alpha$ (b),
$\tau_\alpha$ (c), and $2\tau_\alpha$ (d).
$\tau_\alpha$ is indicated by the arrow.
The wave vector $k$ is chosen as $k=2\pi$.}
\vspace*{-0.5cm}
\label{F4_147_2pi}
\end{figure}

To calculate our three-time correlation function Eq.~(\ref{three_time_correlation}),
we have generated long-time trajectories of
MD simulations for a glass-forming binary mixture composed of soft sphere
components $1$ and $2$.
Particles interact via a soft-core potential
$v_{ab}(r) = \epsilon(\sigma_{ab} /r)^{12}$, where $\sigma_{ab} =
(\sigma_a + \sigma_b)/2$ and $a, b \in \{1, 2\}$.
The interaction is truncated at $r = 3\sigma_1$.
The size and mass ratio are $\sigma_1 / \sigma_2 = 1/1.2$ and $m_1 / m_2
= 1 / 2$, respectively.
The units of length, time, and temperature are $\sigma_1$,
$\tau_0=({m_{1}\sigma_{1}^{2}/\epsilon})^{1/2}$, and $\epsilon/k_B$
in this paper.
Details of the simulations are given in
previous papers~\cite{Yamamoto1998Heterogeneous,
Yamamoto1998Dynamics}.
The systems are composed of $N=N_{1}+N_{2}=1000$ particles
with a fixed density $\rho=N/L^3=0.8$ and a composition $N_1/N=0.5$. 
The corresponding system linear dimension is $L=10.8$.
Simulations were carried out at various temperatures $T=0.772$, $0.473$, $0.352$, 
$0.306$, and $0.289$ with a time step $\Delta t = 0.005$.
Periodic boundary conditions were used in all simulations.
At each temperature, the system was carefully equilibrated in the 
canonical condition, and then,
data were taken in the microcanonical condition.
We have defined the $\alpha$-relaxation time $\tau_{\alpha}$
for each temperature by $F_s(k=2\pi, \tau_{\alpha})=e^{-1}$ for
component $1$ particles, where $k=2\pi$ corresponds to the wave vector of
the first peak of the static structure factor.



In Fig.~\ref{F4_0t2}, we show the two-dimensional plot of
$F_4(k, t_1, t_2, t_3)$ of component $1$ particles  at the zero waiting
time, $t_2=0$, for various temperatures.
$k$ is chosen as $k=2\pi$.
It is demonstrated that the intensity of $F_4(k, t_1, t_2, t_3)$
grows with decreasing $T$, due to correlated motions of heterogeneous dynamics.
This indicates that particles located in
slow (fast) moving regions during
the first time interval $t_1$ 
tend to remain slow (fast) during the second time interval $t_3$.
The line shape of $F_4(k, t_1, t_2, t_3)$ is seen along the diagonal line
$t_1=t_3$, and the time at which $F_4$ has a maximum value is
approximately given by the $\alpha$-relaxation time $\tau_\alpha$.
We here remark that the diagonal part $t_1=t_3$ at $t_2=0$
corresponds to the three-time correlation function used in earlier
studies~\cite{Heuer1997Heterogeneous, Heuer1997Information,
Doliwa1998Cage} to judge whether the relaxation type of the dynamics is
homogeneous or heterogeneous.


\begin{figure}[t]
\includegraphics[width=.25\textwidth]{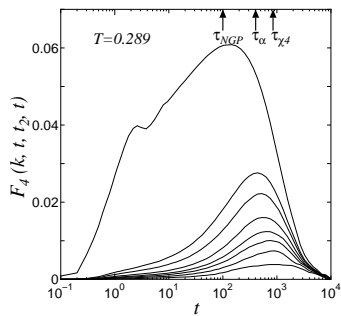}
\vspace*{-0.5cm}
\caption{Diagonal part of the three-time correlation function
 $F_4(k, t, t_2, t)$
 of component $1$ particles
for $T=0.289$ at various waiting times
$t_2=0$, $0.5\tau_\alpha$, $\tau_\alpha$, $2\tau_\alpha$,
 $3\tau_\alpha$, $4\tau_\alpha$, $6\tau_\alpha$, and $10\tau_\alpha$
from top to bottom.
The wave vector $k$ is chosen as $k=2\pi$.
The $\alpha$-relaxation time and the
times at which the non-Gaussian parameter $\alpha_2(t)$ and the dynamic susceptibility
 $\chi_4(k=2\pi, t)$ reach
 their peak values are indicated as $\tau_\alpha$, $\tau_{\rm NGP}$ and
 $\tau_{\chi_4}$, respectively.
}
\vspace*{-0.5cm}
\label{F4_147_t1t3}
\end{figure}

We next investigate how the heterogeneous dynamics evolve with
waiting time $t_2$.
Figure~\ref{F4_147_2pi} shows that the waiting time dependence of $F_4(k,
t_1, t_2, t_3)$ at
the lower temperature $T=0.289$.
In addition, in Fig.~\ref{F4_147_t1t3}, we plot the 
diagonal line along $t_1=t_3$ of $F_4$ at various $t_2$ values.
It is seen that the correlation gradually decays with increasing
the waiting time $t_2$ and tends to become zero for $t_2 \to \infty$.
The presence of motions correlated with the $\alpha$-relaxation time
scale is clearly observed, even for $t_2=10\tau_\alpha$.
Moreover, the off-diagonal part of $F_4$ exists as $t_2$ increases.
This indicates that 
motions between the $\alpha$-relaxation and other relaxations with
different time scales, such
as $\beta$-relaxation, are also coupled, even for large $t_2$ values.
Here, it is seen in Fig.~\ref{F4_147_t1t3} that the peak
position of $F_4$ changes slightly with $t_2$.
For $t_2=0$, the peak appears around the time $\tau_{\rm NGP}$, where
the non-Gaussian parameter $\alpha_2(t) = 3\langle \Delta r^4(t)\rangle /
 5\langle \Delta r^2(t) \rangle^2 -1$ has a peak.
On the other hand, for large $t_2$, we find that the peak of $F_4$
tends to shift to the slower time $\tau_{\chi_4}$, where the non-linear dynamical
susceptibility defined as
$\chi_4(k, t)= N\langle[(1/N) \sum_{j=1}^N
 \delta F_j (\bi{k}, 0, t) ]^2 \rangle$ with $k=2\pi$ has a peak.



\begin{figure}[t]
\includegraphics[width=.25\textwidth]{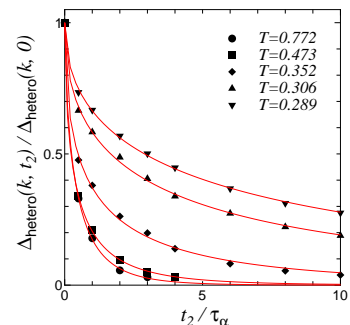}
\vspace*{-0.5cm}
\caption{Waiting time $t_2$ dependence of the integrated three-time
 correlation function $\Delta_{\rm hetero}(k, t_2)/\Delta_{\rm hetero}(k, 0)$
 for various temperatures.
The waiting times are normalized by the $\alpha$ relaxation time
 $\tau_\alpha$ for each temperature.
The wave vector $k$ is chosen as $k=2\pi$.
The solid curve is determined by fitting with the stretched-exponential form
 for each temperature.
}
\vspace*{-0.5cm}
\label{hetero_t2}
\end{figure}

\begin{figure}[t]
\includegraphics[width=.25\textwidth]{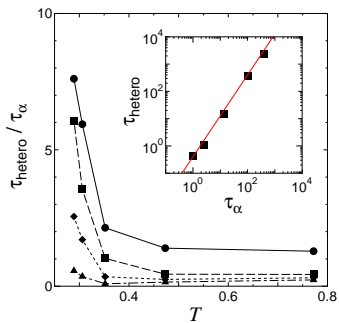}
\vspace*{-0.5cm}
\caption{
Lifetime of heterogeneous dynamics $\tau_{\rm hetero} $ normalized by
 the $\alpha$-relaxation $\tau_\alpha$ 
 versus temperature $T$ for $k=\pi$ (circles), $2\pi$ (squares), $3\pi$
 (diamonds), and $4\pi$ (triangles).
Inset: Relation between $\tau_{\rm hetero}$ and $\tau_\alpha$ for $k=2\pi$.
The straight line with the slope 1.5 is a viewing guide.
}
\vspace*{-0.5cm}
\label{tau_hetero}
\end{figure}

Here, we determine the lifetime of heterogeneous dynamics
$\tau_{\rm hetero}$ and examine
its temperature dependence.
As in the previous studies~\cite{Jung2005Dynamical,
Leonard2005Lifetime}, we define the volume of $F_4(k, t_1, t_2, t_3)$
by integrating over $t_1$ and $t_3$ as
\begin{equation}
\Delta_{\rm hetero}(k, t_2) = \int\int F_4(k, t_1, t_2, t_3) dt_1 dt_3.
\end{equation}
Figure~\ref{hetero_t2} shows $\Delta_{\rm hetero}(k, t_2)/ \Delta_{\rm
hetero}(k, 0)$ with $k=2\pi$
as a function of the waiting time $t_2$ normalized by the
$\alpha$-relaxation time $\tau_\alpha$.
As seen in Fig.~\ref{hetero_t2}, 
$\Delta_{\rm hetero}$ rapidly decays to zero at higher temperatures, and
the time scale is comparable to $\tau_\alpha$.
In contrast, at lower temperatures, the relaxation of $\Delta_{\rm
hetero}$ occurs on a time scale larger than $\tau_\alpha$.
$\Delta_{\rm hetero}(k, t_2)/\Delta_{\rm hetero}(k, 0)$ can be fitted by the
stretched-exponential function $\exp[-(t_2/\tau_{\rm hetero})^c]$, where
$\tau_{\rm hetero}$ can be regarded as 
the lifetime of the heterogeneous dynamics.
We plot $\tau_{\rm hetero}$ for each temperature $T$ and
wave vector $k$ in Fig.~\ref{tau_hetero}.
In Fig.~\ref{tau_hetero}, it is found that $\tau_{\rm hetero}$ becomes
large as the temperature $T$ decreases.
In practice, the lifetime $\tau_{\rm hetero}$ is about $6\tau_\alpha$ with $c \sim 0.5$
at $T=0.289$ and $k=2\pi$.
Furthermore, we confirm that $\tau_{\rm hetero}$ systematically increases with decreasing
$k$, implying that couplings of large scale motions have long time
scales in heterogeneous dynamics.
Such strong deviations and decouplings between $\tau_{\rm hetero}$ and $\tau_\alpha$
with decreasing $T$ have
been observed in experiments~\cite{Wang1999How, Wang2000Lifetime}
and simulations~\cite{Yamamoto1998Heterogeneous, Leonard2005Lifetime, Hedges2007Decoupling}.
A more recent study has demonstrated that the time scale for Fickian
diffusion increases faster than $\tau_\alpha$~\cite{Szamel2006Time},
which would be related to our results.
Another recent study has also mentioned the existence of the slower relaxation time is not
$\tau_\alpha$, but the lifetime of dynamical heterogeneity~\cite{Kawasaki2009Apparent}.

Finally, it is of great interest to examine
the dynamical scaling relation between the time
and length scales of dynamical heterogeneity.
As is mentioned, the four-point correlation function is used to extract
the length scale $\xi$ of the dynamical heterogeneity,
where the fluctuation in the two-point correlation function can be considered
as an order parameter as seen in critical phenomena.
Various dynamical scaling relations have been
proposed~\cite{Yamamoto1998Dynamics, Perera1999Relaxation,
Lacevic2003Spatially},
however the time scale is practically chosen as the
relaxation time of the two-point correlation function, $\tau_\alpha$.
Here, we note that 
the time scale associated with $\xi$ should be
the average lifetime of the fluctuations in the two-point
correlation function, i.e., $\tau_{\rm hetero}$.
From the inset of Fig.~\ref{tau_hetero}, we observe a quantitative power
law behavior between $\tau_{\rm hetero}$ and $\tau_\alpha$ as
$\tau_{\rm hetero} \sim {\tau_\alpha}^\zeta$ with $\zeta=1.5$ for
$k=2\pi$.
A previous study reports that a scaling relation between
$\tau_\alpha$ and $\xi$ as $\tau_\alpha \sim \xi^{z'}$ with $z'=2$ in 
the present binary soft sphere system~\cite{Yamamoto1998Dynamics}.
Note that in Ref.~\cite{Yamamoto1998Dynamics} the correlation length
$\xi$ is estimated in terms of the static structure factors of bond
breakage processes among adjacent particle pairs, which
is essentially the same as the $\xi$ determined by the four-point correlation
function consisting of fluctuations in the local
mobility~\cite{Glotzer2000Spatially, Lacevic2003Spatially,
Berthier2004Time, Toninelli2005Dynamical, Chandler2006Lengthscale}.
Employing this relation, we obtain a new dynamical scaling relation as
$\tau_{\rm hetero} \sim \xi^{z}$ with $z=z' \zeta = 3$, which
characterizes the relaxation processes of the dynamical heterogeneity.

In summary, we have investigated a multi-time correlation
function to quantitatively characterize the time scale of dynamical
heterogeneity.
The two-dimensional plots reveal couplings of particle motions
with various time scales,
and progressive changes in the waiting time allowed
us to extract the time scale of heterogeneous dynamics in
glass-forming liquids.
It is demonstrated that the lifetime of heterogeneous dynamics becomes
rapidly larger than the $\alpha$-relaxation time as the temperature
decreases, which is compatible with recent optical experimental
results~\cite{Wang1999How, Wang2000Lifetime} and computer simulation
studies~\cite{Yamamoto1998Heterogeneous, Leonard2005Lifetime,
Szamel2006Time, Hedges2007Decoupling, Kawasaki2009Apparent}.
We have also studied the wave vector dependence of the lifetime, which is
found to systematically increase for large length scales.
We have furthermore confirmed the dynamical scaling relation between the time
and length scales of dynamical heterogeneity as
$\tau_{\rm hetero} \sim \xi^{z}$ with $z=3$, where both variables are
determined in terms of the four-point correlation function.
It is worth mentioning that
the existence of the slower time scale
$\tau_{\rm hetero}$ than $\tau_\alpha$ may
give the new insights into the characterizing of the violation of the Stokes--Einstein
relation~\cite{Yamamoto1998Heterogeneous, Jung2004Excitation, Hedges2007Decoupling}
and non-Newtonian behaviors in sheared glassy
liquids~\cite{Yamamoto1998Dynamics, Furukawa2009Anisotropic},
as is discussed in Ref.~\cite{Kawasaki2009Apparent}.

This work is partially supported by KAKENHI;
Young Scientists (B)
and Priority Area ``Molecular Theory for Real Systems'', 
the Molecular-Based New Computational Science Program, NINS,
and the Next Generation Super Computing Project, Nanoscience program.
The computations were performed at RCCS, Okazaki, Japan.


\end{document}